\documentclass[epj,twocolumn]{webofc}
\usepackage[varg]{txfonts}   
\usepackage{subfigure}
\usepackage{amssymb,amsmath}

\graphicspath{{fig/}}
\woctitle{QUARKS 2016}
\begin{document}
\title{Recent results from the Belle experiment}
%
%

\author{\firstname{Dmitri} \lastname{Liventsev}\inst{1,2}\fnsep\thanks{\email{dmitri.liventsev@kek.jp}}\fnsep\thanks{On behalf of the Belle collaboration}
}

\institute{Virginia Polytechnic Institute and State University, Blacksburg, Virginia 24061
\and
           High Energy Accelerator Research Organization (KEK), Tsukuba 305-0801
          }

\abstract{We review recent results obtained using the data recorded with
  the Belle detector at the KEKB asymmetric-energy $e^+ e^-$ collider in
  KEK, Japan.}
\maketitle
\section{Introduction}
\label{intro}

The Belle detector is a large-solid-angle magnetic spectrometer that
consists of a silicon vertex detector (SVD), a 50-layer central drift
chamber (CDC), an array of aerogel threshold Cherenkov counters (ACC), a
barrel-like arrangement of time-of-flight scintillation counters (TOF),
and an electromagnetic calorimeter comprised of CsI(Tl) crystals (ECL)
located inside a super-conducting solenoid coil that provides a 1.5~T
magnetic field. An iron flux-return located outside of the coil is
instrumented to detect $K_L^0$ mesons and to identify muons (KLM). The
detector is described in detail elsewhere~\cite{BelleDetektor}.

The Belle experiment successfully operated for more than a decade until
2010 at the asymmetric-energy $e^+e^-$ collider KEKB~\cite{KEKB} in
various $\Upsilon(nS)$ resonances, having collected a world-record
sample of data over $1\,\mathrm{ab}^{-1}$.

Here we present a review of recent results from Belle based on its full
statistics.

\section{Angular Analysis of $B^0 \to K^\ast(892)^0 \ell^+ \ell^-$}
\label{sec-1-kll}

Rare decays of $B$ mesons are an ideal probe to search beyond the
Standard Model (SM) of particle physics, since contributions from new
particles lead to effects that are of similar size as the SM
predictions. The rare decay $B^0 \to K^\ast(892)^0 \ell^+ \ell^-$, where
$\ell^+\ell^-$ is either $e^+e^-$ or $\mu^+\mu^-$, involves the quark
transition $b\to s \ell^+ \ell^-$, a flavor changing neutral current
that is forbidden at tree level in the SM. Higher order SM processes
such as penguin or $W^+W^-$ box diagrams allow for such transitions,
leading to branching fractions of less than one in a million. Various
extensions to the SM predict contributions from new physics, which can
interfere with the SM amplitudes and lead to enhanced or suppressed
branching fractions or modified angular distributions of the decay
products.

Belle presented an angular analysis~\cite{kll}, using the decay modes
$B^0 \to K^\ast(892)^0 e^+ e^-$ and $B^0 \to K^\ast(892)^0 \mu^+ \mu^-$,
in a data sample recorded with the Belle detector. The LHCb
collaboration reported a discrepancy in the angular distribution of the
decay $B^0 \to K^\ast(892)^0 \mu^+ \mu^-$, corresponding to a
$3.4\sigma$ deviation from the SM prediction \cite{lhcb2}. In contrast
to the LHCb measurement the di-electron channel is also used in this
analysis.

$K^\ast$ candidates are formed in the channel $K^{\ast0}\to K^+\pi^-$
and are combined with oppositely charged lepton pairs to form $B$ meson
candidates.  The large combinatoric background is suppressed by applying
requirements on kinematic variables.  Two independent variables can be
constructed using constraints that in $\Upsilon(4S)$ decays $B$ mesons
are produced pairwise and each carries half the center--of--mass (CM)
frame beam energy, $E_{\textrm{Beam}}$.  These variables are the beam
constrained mass, $M_\textrm{bc}$, and the energy difference, $\Delta
E$, in which signal features a distinct distribution that can
discriminate against background.  The variables are defined in the
$\Upsilon(4S)$ rest frame as
\begin{align}
        M_\textrm{bc} & \equiv \sqrt{E^2_{\mathrm{Beam}}/c^4 -|\vec p_B|^2/c^2}~\mathrm{and} \\
        \Delta E & \equiv E_B - E_{\mathrm{Beam}},\label{eq:deltaE}
\end{align}
where $E_B$ and $|\vec p_B|$ are the energy and momentum of the
reconstructed candidate, respectively.  Correctly reconstructed
candidates are located around the nominal $B$ mass in $M_\textrm{bc}$
and feature $\Delta E$ of around zero.  Candidates are selected
satisfying $ 5.22~ < M_\mathrm{bc} <~5.3~\mathrm{GeV}/c^2 $ and $ -0.10
\ (-0.05)~ <\Delta E < ~0.05~\mathrm{GeV}$ for $\ell=e$ ($\ell=\mu$).

Large irreducible background contributions arise from charmonium decays
$B\to K^{(\ast)} J/\psi$ and $B\to K^{(\ast)} \psi(2S)$, in which the
$c\bar c$ state decays into two leptons.  These decays have the same
signature as the desired signal and are vetoed with the following
requirements on $q^2=M_{\ell^+\ell^-}$, the invariant mass of the lepton
pair:
$-0.25~\mathrm{GeV}/c^2 <  M_{ee(\gamma)} - m_{J/\psi}< 0.08~\mathrm{GeV}/c^2$,
$-0.15~\mathrm{GeV}/c^2 <  M_{\mu\mu} - m_{J/\psi}< 0.08~\mathrm{GeV}/c^2$,
$-0.20~\mathrm{GeV}/c^2 <  M_{ee(\gamma)} - m_{\psi(2S)}< 0.08~\mathrm{GeV}/c^2$ and
$-0.10~\mathrm{GeV}/c^2 <  M_{\mu\mu} - m_{\psi(2S)}< 0.08~\mathrm{GeV}/c^2$.


For the angular analysis the number of signal events $n_\mathrm{sig}$
and background events $n_{\mathrm{bkg}}$ in the signal region
$M_\textrm{bc} >5.27~\mathrm{GeV}/c^2$ are obtained by a fit to
$M_\textrm{bc}$ in bins of $q^2$. The extracted yields and the
definition of the bin ranges are presented in Table
\ref{tab:signal_yields}.
\begin{table}
  \centering
  \caption{Fitted yields and statistical error for signal ($n_{\rm sig}$) and background ($n_{\rm bkg}$) events in the binning of $q^2$ for both the combined electron and muon channel.}
  \label{tab:signal_yields}
  \begin{tabular}{cccc}
      \hline \hline
      Bin &   $q^2$ range in $\textrm{GeV}^2/c^4$ &  $n_\textrm{sig}$ &  $n_\textrm{bkg}$  \\ 
      \hline
      0 & $1.00 - 6.00$  &  $49.5\pm 8.4$  &  $30.3\pm 5.5$  \\
      \hline 
      1 &$0.10 - 4.00$  &  $30.9\pm 7.4$  &  $26.4\pm 5.1$  \\
      2 &$4.00 - 8.00$  &  $49.8\pm 9.3$  &  $35.6\pm 6.0$ \\
      3 &$10.09 - 12.90$  &  $39.6\pm 8.0$  &  $19.3\pm 4.4$  \\
      4 &$14.18 - 19.00$  &  $56.5\pm 8.7$  &  $16.0\pm 4.0$   \\
      \hline \hline
  \end{tabular}
\end{table}

We perform an angular analysis of $B^0 \to K^\ast(892)^0 \ell^+ \ell^-$
including the electron and muon modes. The decay is kinematically
described by three angles $\theta_\ell$, $\theta_K$ and $\phi$ and the
invariant mass squared of the lepton pair $q^2$. Definitions of the
angles and the full angular distribution follow Ref.~\cite{lhcbafb}.
The binning in $q^2$ is
detailed in Table~\ref{tab:signal_yields} together with the measured
signal and background yields. Uncovered regions in the $q^2$ spectrum
arise from vetoes against backgrounds of the charmonium resonances
$J/\psi \to \ell^+\ell^-$ and $\psi(2S) \to \ell^+\ell^-$ and vetos
against $\pi^0$ Dalitz decays and photon conversion.

The observables $P_{i=4,5,6,8}'$, introduced in
Ref.~\cite{DescotesGenon1} are functions of Wilson coefficients,
containing information about the short-distance effects and can be
affected by new physics and are considered to be largely free from
form-factor uncertainties \cite{DescotesGenon2}. The statistics in this
analysis are not sufficient to perform an eight-dimensional fit,
therefore a folding technique is used explained in more detail in
Refs.~\cite{lhcb1}.

The signal and background fractions are derived from a fit to 
beforehand, where the yields are listed in Table
\ref{tab:signal_yields}. The $M_\textrm{bc}$ variable is split into a signal
(upper) and sideband (lower) region at $5.27~\mathrm{GeV}/c^2$. In the
second step the shape of the background for the angular observables is
estimated on the $M_\textrm{bc}$ sideband. This is possible as the angular
observables have shown to be uncorrelated to $M_\textrm{bc}$ in the background
sample.
        
All observables $P_{i=4,5,6,8}'$ are extracted from the data in the
signal region using three-dimensional unbinned maximum likelihood fits
in four bins of $q^2$ and the additional zeroth bin using the folded
signal PDF, fixed background shapes and a fixed number of signal
events. Each $P_{i=4,5,6,8}'$ is fitted with the $K^\ast$ longitudinal
polarization $F_L$ and the transverse polarization asymmetry
$A_T^{(2)}$. Counting also the zeroth bin, which exhibits overlap with
the range of the first and second bin, 20 measurements are performed.

\begin{figure}[h]
  \centering
  \includegraphics[width=0.46\textwidth,clip=true]{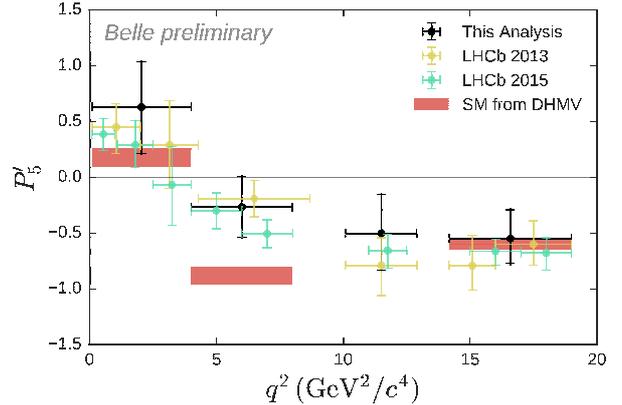}
  \caption{Result for the $P_5'$ observable compared to SM predictions
    from various sources. Results from LHCb \cite{lhcb1,lhcb2} are shown
    for comparison.}
  \label{fig:kll-res}
\end{figure}

The measurements are compared with SM predictions based upon different
theoretical calculations. The result for $P_5'$ is shown in
Fig.~\ref{fig:kll-res} together with available SM prediction and LHCb
measurements. A deviation with respect to the SM prediction is observed
with a significance of $2.1\sigma$ in the $q^2$ range $4.0 < q^2 < 8.0
~\mathrm{GeV}^2/c^4$.
This deviation is into the same direction and in the same $q^2$ region
where the LHCb collaboration reported the so-called $P_5'$ anomaly
\cite{lhcb1,lhcb2}.


\section{Measurement of the branching ratio of $\bar{B}^0 \rightarrow D^{*+} \tau^- \bar{\nu}_{\tau}$}
\label{sec-2-dtaunu}

Semitauonic $B$ meson decays of the type $b \rightarrow c \tau
\nu_{\tau}$ are sensitive probes to search for physics beyond the
Standard Model (SM). Charged Higgs bosons, which appear in
supersymmetry and other models with at least two Higgs doublets, may
contribute to the decay to due to large mass of the $\tau$ lepton and
induce measurable effects in the branching fraction.  Similarly,
leptoquarks, which carry both baryon number and lepton number, may also
contribute to this process.  The ratio of branching fractions
\begin{eqnarray}
{\cal R}(D^{(*)}) = \frac{{\cal B}(\bar{B} \rightarrow D^{(*)} \tau^- \bar{\nu}_{\tau})}{{\cal B}(\bar{B} \rightarrow D^{(*)} \ell^- \bar{\nu}_{\ell})} \hspace{0.8em}(\ell = e,\mu),
\end{eqnarray}
is typically used instead of the absolute branching fraction of $\bar{B}
\rightarrow D^{(*)+} \tau^- \bar{\nu}_{\tau}$, to reduce several
systematic uncertainties such as those on the experimental efficiency,
the CKM matrix elements $|V_{cb}|$, and on the form factors. The SM
calculations on these ratios predict ${\cal R}(D^*) = 0.252 \pm
0.003$~\cite{SM_PREDICTION_2} and ${\cal R}(D) = 0.297 \pm
0.017$~\cite{SM_PREDICTION_1,BELLE_HAD_NEW} with precision of better
than 2\% and 6\% for ${\cal R}(D^*)$ and ${\cal R}(D)$, respectively.
Exclusive semitauonic $B$ decays were first observed by the Belle
Collaboration \cite{BELLE_INCLUSIVE_OBSERVATION}, with subsequent
studies reported by Belle \cite{BELLE_INCLUSIVE,BELLE_HAD_NEW},
\mbox{\sl B\hspace{-0.4em} {\small\sl A}\hspace{-0.37em} \sl
  B\hspace{-0.4em} {\small\sl A\hspace{-0.02em}R}} \cite{BABAR_HAD_NEW},
and LHCb \cite{LHCB_RESULT} Collaborations. All results are consistent
with each other, and the average values of
Refs.~\cite{BELLE_HAD_NEW,BABAR_HAD_NEW,LHCB_RESULT} have been found to
be ${\cal R}(D^*) = 0.322 \pm 0.018 \pm 0.012$ and ${\cal R}(D) = 0.391
\pm 0.041 \pm 0.028$ \cite{HFAG}, which exceed the SM predictions for
${\cal R}(D^*)$ and ${\cal R}(D)$ by $3.0 \sigma$ and $1.7 \sigma$,
respectively.  The combined analysis of ${\cal R}(D^*)$ and ${\cal
  R}(D)$, taking into account measurement correlations, finds that the
deviation is $3.9\sigma$ from the SM prediction.

In the paper~\cite{bdtaunu} Belle reported the first measurement of
${\cal R}(D^*)$ using the semileptonic tagging method. Signal
$B^0\bar{B}^0$ events are reconstructed in modes where one $B$ decays
semi-tauonically $\bar{B}^0 \rightarrow D^{*+} \tau^- \bar{\nu}_{\tau}$
where $\tau^- \rightarrow \ell^- \bar{\nu}_{\ell} \nu_{\tau}$, (referred
to hereafter as $B_{\rm sig}$) and the the other $B$ decays in a
semileptonic channel $\bar{B}^0 \rightarrow D^{*+} \ell^-
\bar{\nu}_{\ell}$ (referred to hereafter as $B_{\rm tag}$). To
reconstruct normalization $B^0\bar{B}^0$ events, which correspond to the
denominator in ${\cal R}(D^*)$, both $B$ mesons are reconstructed
decaying to semileptonic decay modes $D^{*\pm} \ell^{\mp}
\bar{\nu}_{\ell}$.

To tag semileptonic $B$ decays, we combine $D^{*+}$ meson and lepton
candidates of opposite electric charge and calculate the cosine of the
angle between the momentum of the $B$ meson and the $D^* \ell$ system in
the $\Upsilon(4S)$ rest frame, under the assumption that only one
massless particle is not reconstructed:
\begin{eqnarray}
\cos \theta_{B \mathchar`-  D^* \ell} \equiv
\frac
{2E_{\rm beam} E_{D^* \ell} - m_B^2 - M_{D^* \ell}^2}
{2 |\vec{p}_B| \cdot |\vec{p}_{D^* \ell}|},
\label{eq:cos_bdstrl}
\end{eqnarray}
where $E_{\rm beam}$ is the energy of the beam, and $E_{D^* \ell}$,
$\vec{p}_{D^* \ell}$ and $M_{D^* \ell}$ are the energy, momentum, and
mass of the $D^* \ell$ system, respectively.  The variable $m_B$ is the
nominal $B$ meson mass~\cite{pdg}, and $\vec{p}_B$ is the nominal $B$
meson momentum.  All variables are defined in the $\Upsilon(4S)$ rest
frame.  Correctly reconstructed $B$ candidates in the tag and
normalization mode $D^* \ell \nu_{\ell}$ are expected to have a value of
$\cos \theta_{B \mathchar`- D^* \ell}$ between $-1$ and $+1$.  On the
other hand, correctly reconstructed $B$ candidates in the signal decay
mode $D^* \tau \nu_{\tau}$ or falsely reconstructed $B$ candidates would
tend to have values of $\cos \theta_{B \mathchar`- D^* \ell}$ below the
physical region due to contributions from additional particles and a
large negative correlation with missing mass squared, $M_{\rm miss}^2 =
(2E_{\rm beam} - \sum_i E_i)^2/c^4 - |\sum_i \vec{p}_i|^2/c^2$, where
$(\vec{p}_i, E_i)$ is four-momentum of the particles in the
$\Upsilon(4S)$ rest frame.

In each event we require two tagged $B$ candidates that are opposite in
flavor. Signal events may have the same flavor due to the $B \bar{B}$
mixing, however we veto such events as they lead to ambiguous $D^* \ell$
pair assignment and larger combinatorial background. We require that at
most one $B$ meson is reconstructed in a $D^+$ mode, in order to avoid
large background from fake neutral pions when forming $D^*$ candidates.
In each signal event we assign the candidate with the lowest value of
$\cos \theta_{B \mathchar`- D^* \ell}$ (referred to hereafter as $\cos
\theta_{B \mathchar`- D^* \ell}^{\rm sig}$) as $B_{\rm sig}$.

To separate reconstructed signal and normalization events, we employ a
neural network approach. The variables used as inputs to the network are
(i) $\cos \theta^{\rm sig}_{B \mathchar`- D^* \ell}$, (ii) missing mass
squared, $M_{\rm miss}^2$, and (iii) visible energy $E_{\rm vis} =
\sum_i E_i$, where $E_i$ is energies of the particles in the
$\Upsilon(4S)$ rest frame.
To separate signal and normalization events from background processes,
we use the extra energy, $E_{\rm ECL}$, which is defined as the sum of
the energies of neutral clusters detected in the ECL that are not
associated with reconstructed particles. 

\begin{figure}[h]
  \centering
  \includegraphics[width=0.46\textwidth]{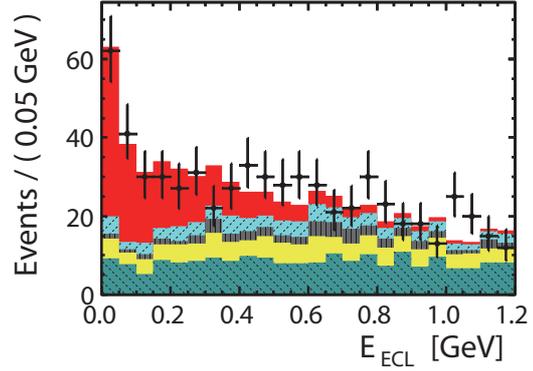}
  \caption{Projections of the fit results with data points overlaid of
    $E_{\rm ECL}$ distribution with signal-enhanced $\mathit{NN}$ region
    ($\mathit{NN} > 0.8$.  The background categories are described in
    detail in the text, where ``others'' refers to predominantly $B
    \rightarrow X_c D^*$ decays.}
  \label{fig:dtaunu-result_fit}
\end{figure}

We extract the signal and normalization yields using a two-dimensional
extended maximum-likelihood fit in $\mathit{NN}$ and $E_{\rm ECL}$.
The projection of the fitted distributions are shown in
Figure~\ref{fig:dtaunu-result_fit}. The yields of signal and normalization
events are measured to be $231 \pm 23({\rm stat})$ and $2800 \pm 57({\rm
  stat})$, respectively. The ratio ${\cal R}(D^*)$ is therefore found to
be
\begin{eqnarray}
{\cal R}(D^*) &=& 0.302 \pm 0.030 \pm 0.011,
\end{eqnarray}
where the first and second errors correspond to statistical and
systematic uncertainties, respectively.

We calculate the statistical significance of the signal as $\sqrt{-2\ln
  ({\cal L}_0/{\cal L}_{\rm max})}$, where ${\cal L}_{\rm max}$ and
${\cal L}_0$ are the maximum likelihood and the likelihood obtained
assuming zero signal yield, respectively. We obtain a statistical
significance of $13.8\sigma$. We also estimate the compatibility of the
measured value of ${\cal R}(D^*)$ and the SM prediction. The effect of
systematic uncertainties are included by convolving the likelihood
function with a Gaussian distribution. We obtain that our result is
larger than the SM prediction by $1.6\sigma$.

We investigated the compatibility of the data samples with type II
two-Higgs-doublet model (2HDM) and leptoquark models.
We find our data is compatible with the SM and type II 2HDM
with $\tan \beta / m_{H^+} = 0.7$ GeV$^{-1}$, while the $R_2$ type
leptoquark model with $C_T = +0.36$ is disfavored.


\section{Observation of the decay $B^0_s \to K^0 \bar{K}^0$}
\label{sec-3-bskk}

The two-body decays $B^0_s \to h^+h'^-$, where
$h^{\scriptscriptstyle(}\kern-1pt{}'\kern-1pt{}^{\scriptscriptstyle)}$
is either a pion or kaon, have now all been observed~\cite{pdg}. In
contrast, the neutral-daughter decays $B^0_s\to h^0h'^0$ have yet to be
observed. The decay $B^0_s\to K^0\bar{K}^0$ is of particular interest
because the branching fraction is predicted to be relatively large. In
the Standard model, the decay proceeds mainly via a $b \to s$ loop (or
``penguin'') transition and the branching fraction is predicted to be in
the range $(16-27)\times10^{-6}$~\cite{SM-branching}. The presence of
non-SM particles or couplings could enhance this
value~\cite{Chang:2013hba}. It has been pointed out that $CP$
asymmetries in $B^0_s\to K^0\bar{K}^0$ decays are promising observables
in which to search for new physics~\cite{susy}.

The current upper limit on the branching fraction, $\mathcal{B}(B^0_s\to
K^0\bar{K}^0)<6.6\times 10^{-5}$ at 90\% confidence level, was set by
the Belle Collaboration using $23.6~{\rm fb^{-1}}$ of data recorded at
the $\Upsilon(5S)$ resonance~\cite{Peng:2010ze}. In paper~\cite{bskk} Belle
updates this result using the full data set of $121.4~{\rm fb^{-1}}$
recorded at the~$\Upsilon(5S)$. The analysis presented here uses
improved tracking, $K^0$ reconstruction, and continuum suppression
algorithms. The data set corresponds to $(6.53\pm 0.66)\times10^6$
$B^0_s\bar{B}^0_s$ pairs~\cite{Oswald:2015dma} produced in three
$\Upsilon(5S)$ decay channels: $B^0_s\bar{B}^0_s$, $B^{*0}_s
\bar{B}^0_s$ or $B^0_s\bar{B}^{*0}_s$, and $B^{*0}_0\bar{B}^{*0}_s$. The
latter two channels dominate, with production fractions of
$f_{B^{*0}_s\bar{B}^0_s}=(7.3\pm1.4)\%$ and
$f_{B^{*0}_s\bar{B}^{*0}_s}=(87.0\pm1.7)$\%~\cite{Esen:2012yz}.

Candidate $K^0$ mesons are reconstructed via the decay $K_s \to
\pi^+\pi^-$ using a neural network (NN) technique. To identify $B_s^0
\to K_s K_s$ candidates, we define two variables: the
beam-energy-constrained mass $M_{\rm bc}=\sqrt{E^2_{\rm
    beam}-|\vec{p}^{}_{B}|^2c^2}/c^2$; and the energy difference $\Delta
E=E_{B}-E_{\rm beam}$, where $E_{\rm beam}$ is the beam energy and $E_B$
and $\vec{p}^{}_{B}$ are the energy and momentum, respectively, of the
$B_s^0$ candidate. These quantities are evaluated in the $e^+e^-$
center-of-mass frame. We require that events satisfy $M_{\rm bc} >
5.34$~GeV/$c^2$\ and $-0.20~{\rm GeV} < \Delta E< 0.10~{\rm GeV}$. To
suppress background arising from continuum $e^+e^-\to
q\bar{q}~(q=u,d,s,c)$ production, we use a second NN that distinguishes
jetlike continuum events from more spherical
$B_s^{(*)0}\bar{B}_s^{(*)0}$ events. The NN has a single output variable
($C_{\rm NN}$) that ranges from $-1$ for backgroundlike events to $+1$
for signal-like events. We require $C_{\rm NN}>-0.1$, which rejects
approximately 85\% of $q\bar{q}$ background while retaining 83\% of
signal decays. We subsequently translate $C_{\rm NN}$ to a new variable
\begin{equation}
C'_{\rm NN} = \ln\left(\frac{C_{\rm NN}-C^{\rm min}_{\rm NN}}
{C^{\rm max}_{\rm NN}-C_{\rm NN}}\right),
\end{equation} 
where $C^{\rm min}_{\rm NN}= -0.1$ and $C_{\rm NN}^{\rm max}$ is the
maximum value of $C_{\rm NN}$ obtained from a large sample of signal MC
decays. The distribution of $C'_{\rm NN}$ is well modeled by a Gaussian
function. Backgrounds arising from other $B^0_s$ and non-$B^0_s$
decays were studied using MC simulation and found to be negligible.

After applying all selection criteria, approximately 1.0\% of events
have multiple $B^0_s$ candidates. For these events, we retain the
candidate having the smallest value of $\chi^2$ obtained from the
deviations of the reconstructed $K_s$ masses from their nominal
values~\cite{pdg}. According to MC simulation, this criterion selects
the correct $B^0_s$ candidate $>99$\% of the time.

We measure the signal yield by performing an unbinned extended maximum
likelihood fit to the variables $M_{\rm bc}$, $\Delta E$, and
$C^{\prime}_{\rm NN}$. 

\begin{figure*}
  \includegraphics[width=0.32\textwidth]{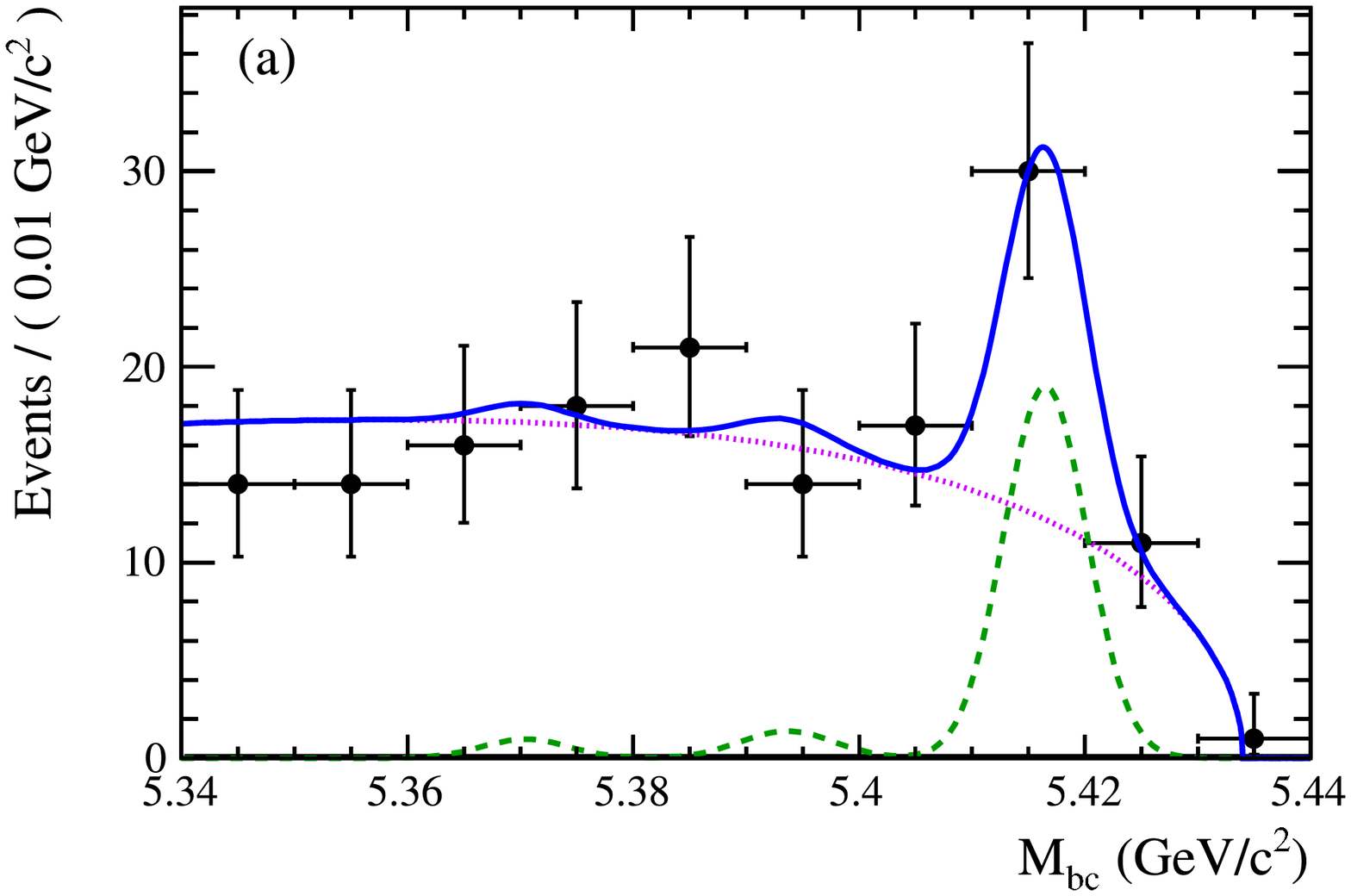}
  \includegraphics[width=0.32\textwidth]{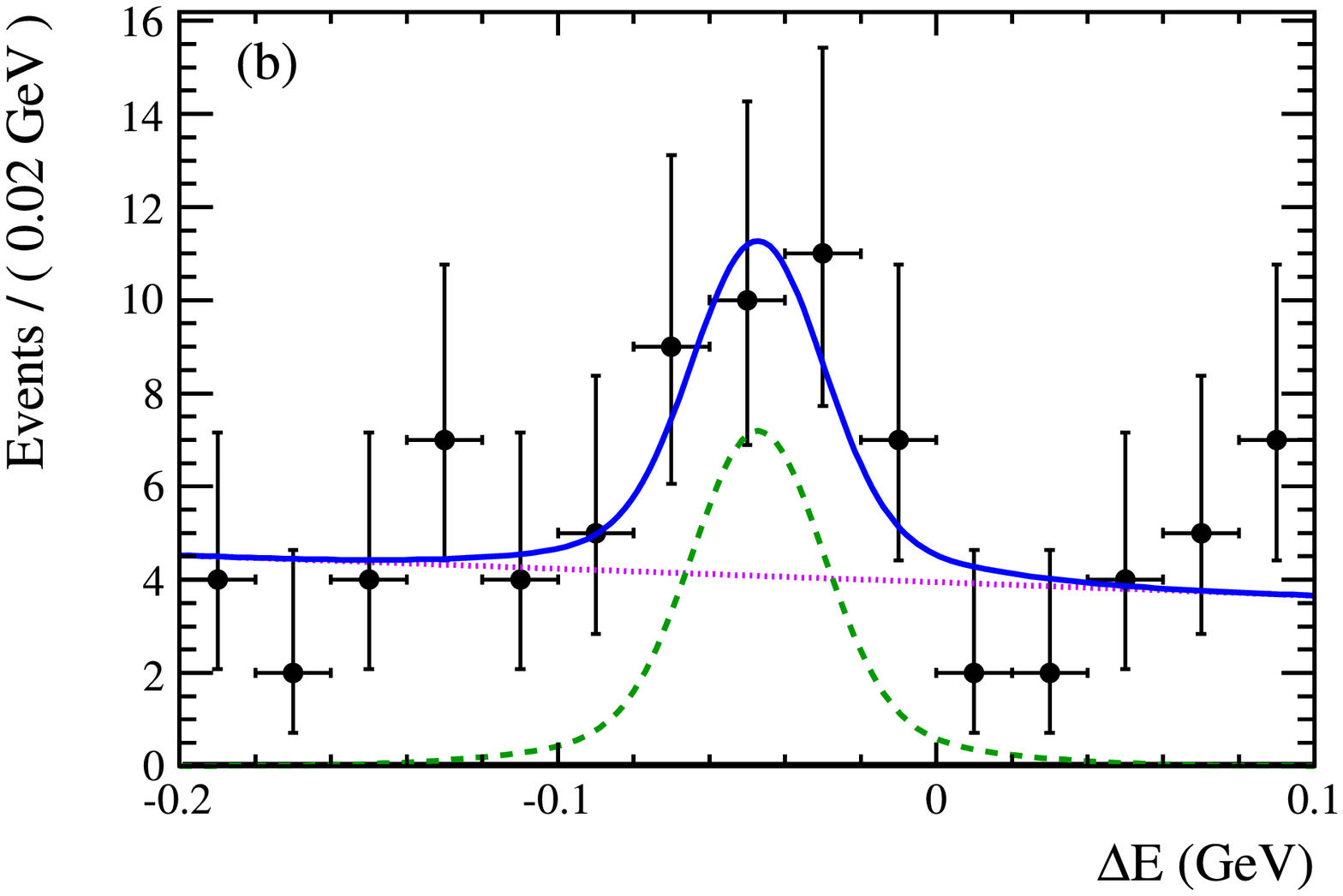}
  \includegraphics[width=0.32\textwidth]{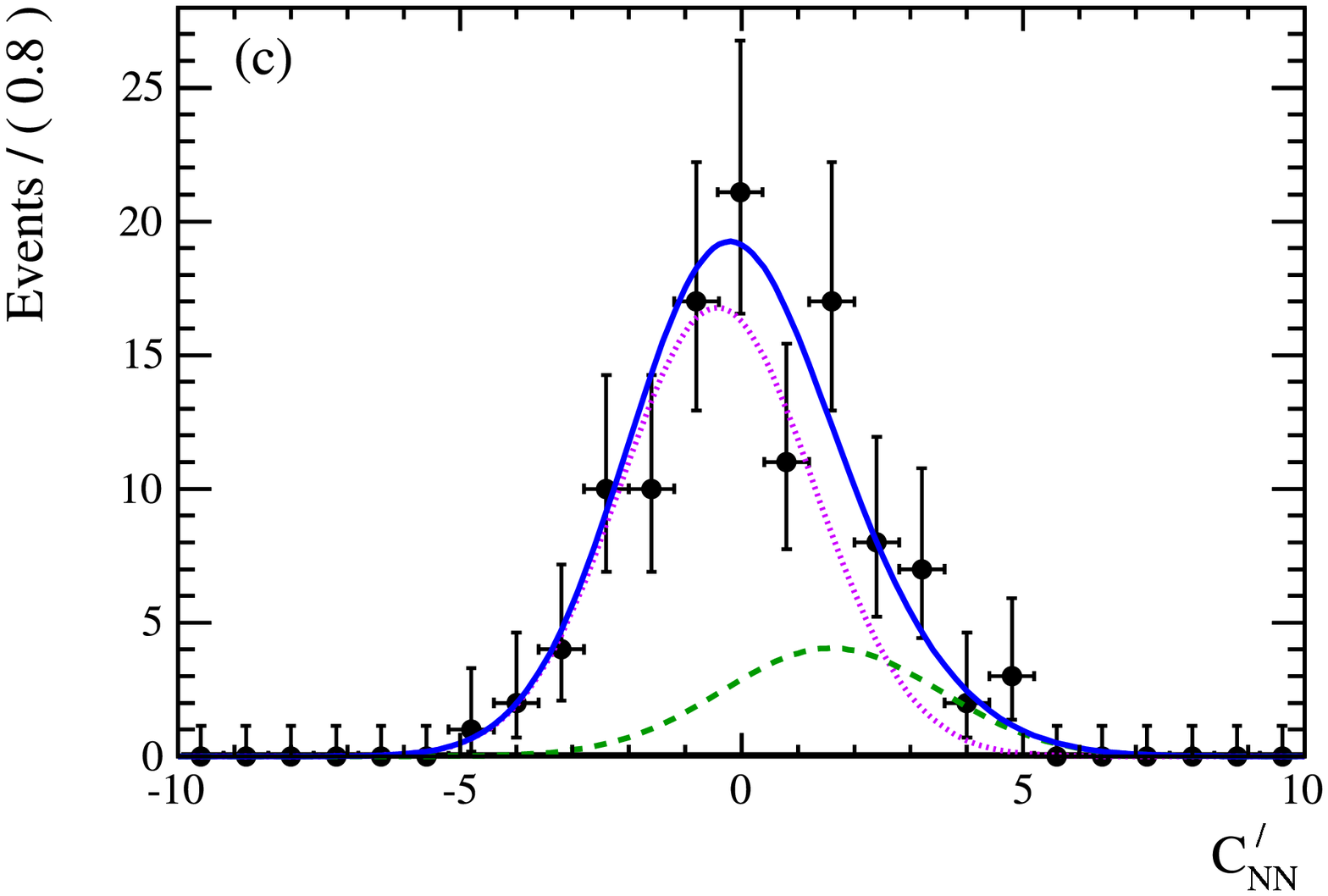}
\caption{\small Projections of the 3D fit to the real data: (a) $M_{\rm
    bc}$ in $-0.11~{\rm GeV} <\Delta E < 0.02~{\rm GeV}$ and
  $C^{\prime}_{\rm NN}>0.5$; (b) $\Delta E$ in $5.405~{\rm GeV}/c^{2}
  <M_{\rm bc}< 5.427~{\rm GeV}/c^{2}$ and $C^{\prime}_{\rm NN}>0.5$; and
  (c) $C^{\prime}_{\rm NN}$ in $5.405~{\rm GeV}/c^{2} <M_{\rm bc}<
  5.427~{\rm GeV}/c^{2}$ and $-0.11~{\rm GeV} <\Delta E < 0.02~{\rm
    GeV}$. The points with error bars are data, the (green) dashed
  curves show the signal, (magenta) dotted curves show the continuum
  background, and (blue) solid curves show the total. The $\chi^2/{\rm
    (number~ of~ bins)}$ values of these fit projections are 0.30, 0.43,
  and 0.26, respectively, which indicate that the fit gives a good
  description of the data. The three peaks in $M_{\rm bc}$ arise from
  $\Upsilon(5S)\to B^0_s\bar{B}^0_s, B^{*0}_s\bar{B}^0_s +
  B^0_s\bar{B}^{*0}_s$, and $B^{*0}_s B^0_s\bar{B}^{*0}_s$ decays.}
\label{fig:bskk-fig2}
\end{figure*}

The results of the fit are $29.0\,^{+8.5}_{-7.6}$ signal events and
$1095.0\,^{+33.9}_{-33.4}$ continuum background events. Projections of
the fit are shown in Fig.~\ref{fig:bskk-fig2}. The branching fraction
is calculated via
\begin{eqnarray}
\mathcal{B}(B^0_s \to K^0 \bar{K}^0) & = &
\frac{Y_{s}}{2 N_{B^0_s\bar{B}^0_s}(0.50)
\mathcal{B}^2_{K^0} \varepsilon},
\end{eqnarray}
where $Y_{s}$ is the fitted signal yield; $N_{B^0_s\bar{B}^0_s}=(6.53\pm
0.66)\times10^6$ is the number of $B^0_s\bar{B}^0_s$ events;
$\mathcal{B}_{K^0}=(69.20\pm0.05)\%$ is the branching fraction for
$K_s\to\pi^+\pi^-$~\cite{pdg}; and $\varepsilon=(46.3\pm 0.1)\%$ is the
signal efficiency as determined from MC simulation. The factor 0.50
accounts for the 50\% probability for $K^0\bar{K}^0\to K_s K_s$ (since
$K^0\bar{K}^0$ is $CP$ even). Inserting these values gives
$\mathcal{B}(B^0_s\to K^0 \bar{K}^0) =
(19.6\,^{+5.8}_{-5.1}\,\pm1.0\,\pm2.0)\times10^{-6}$, where the first
uncertainty is statistical, the second is systematic, and the third
reflects the uncertainty due to the total number of $B^0_s\bar{B}^0_s$
pairs. This value is in good agreement with the SM
predictions~\cite{SM-branching}, and it implies that the Belle~II
experiment~\cite{TDR} will reconstruct over 1000 of these decays. Such a
sample would allow for a much higher sensitivity search for new physics
in this $b\to s$ penguin-dominated decay.

The signal significance is calculated as
$\sqrt{-2\ln(\mathcal{L}_0/\mathcal{L}_{\rm max})}$, where
$\mathcal{L}_0$ is the likelihood value when the signal yield is fixed
to zero, and $\mathcal{L}_{\rm max}$ is the likelihood value of the
nominal fit. We include systematic uncertainties in the significance by
convolving the likelihood function with a Gaussian function whose width
is equal to that part of the systematic uncertainty that affects the
signal yield. We obtain a signal significance of 5.1 standard
deviations; thus, our measurement constitutes the first observation of
this decay.

\section{Study of $e^+e^- \to B^{(*)}\bar{B}^{(*)}\pi^\pm$ at $\sqrt{s}=10.866$~GeV}
\label{sec-4-zb}

Two new charged bottomonium-like resonances, $Z_b(10610)$ and
$Z_b(10650)$, have been observed recently by the Belle Collaboration in
$e^+e^-\to\Upsilon(n{\rm S})\pi^+\pi^-$, $n=1,2,3$ and $e^+e^-\to h_b(m{\rm
  P})\pi^+\pi^-$, $m=1,2$~\cite{y5s2ypp,y5s2ypp2}. Analysis of the quark
composition of the initial and final states reveals that these hadronic
objects have an exotic nature: $Z_b$ should be comprised of (at least)
four quarks including a $b\bar{b}$ pair. Several models~\cite{zbmodels}
have been proposed to describe the internal structure of these
states. In Ref.~\cite{molec}, it was suggested that $Z_b(10610)$ and
$Z_b(10650)$ states might be loosely bound $B\bar{B}^*$ and
$B^*\bar{B}^*$ systems, respectively. If so, it is natural to expect the
$Z_b$ states to decay to final states with $B^{(*)}$ mesons at
substantial rates.

Evidence for the three-body $\Upsilon(10860)\to B\bar{B}^*\pi$ decay has
been reported previously by Belle, based on a data sample of
$23.6$~fb$^{-1}$~\cite{drutskoj}. In the analysis~\cite{zb} Belle uses a
data sample with an integrated luminosity of $121.4$~fb$^{-1}$ collected
near the peak of the $\Upsilon(10860)$ resonance ($\sqrt{s}=10.866$~GeV)
with the Belle detector. Note that we reconstruct only three-body
$B^{(*)}\bar{B}^{(*)}\pi$ combinations with a charged primary pion. For
brevity, we adopt the following notations: the set of
$B^+\bar{B}^0\pi^-$ and $B^-B^0\pi^+$ final states is referred to as
$BB\pi$; the set of $B^+\bar{B}^{*0}\pi^-$, $B^-B^{*0}\pi^+$,
$B^0B^{*-}\pi^+$ and $\bar{B}^0B^{*+}\pi^-$ final states is referred to
as $BB^*\pi$; and the set of $B^{*+}\bar{B}^{*0}\pi^-$ and
$B^{*-}B^{*0}\pi^+$ final states is denoted as $B^*B^*\pi$.

\begin{figure}[h]
  \includegraphics[width=0.245\textwidth]{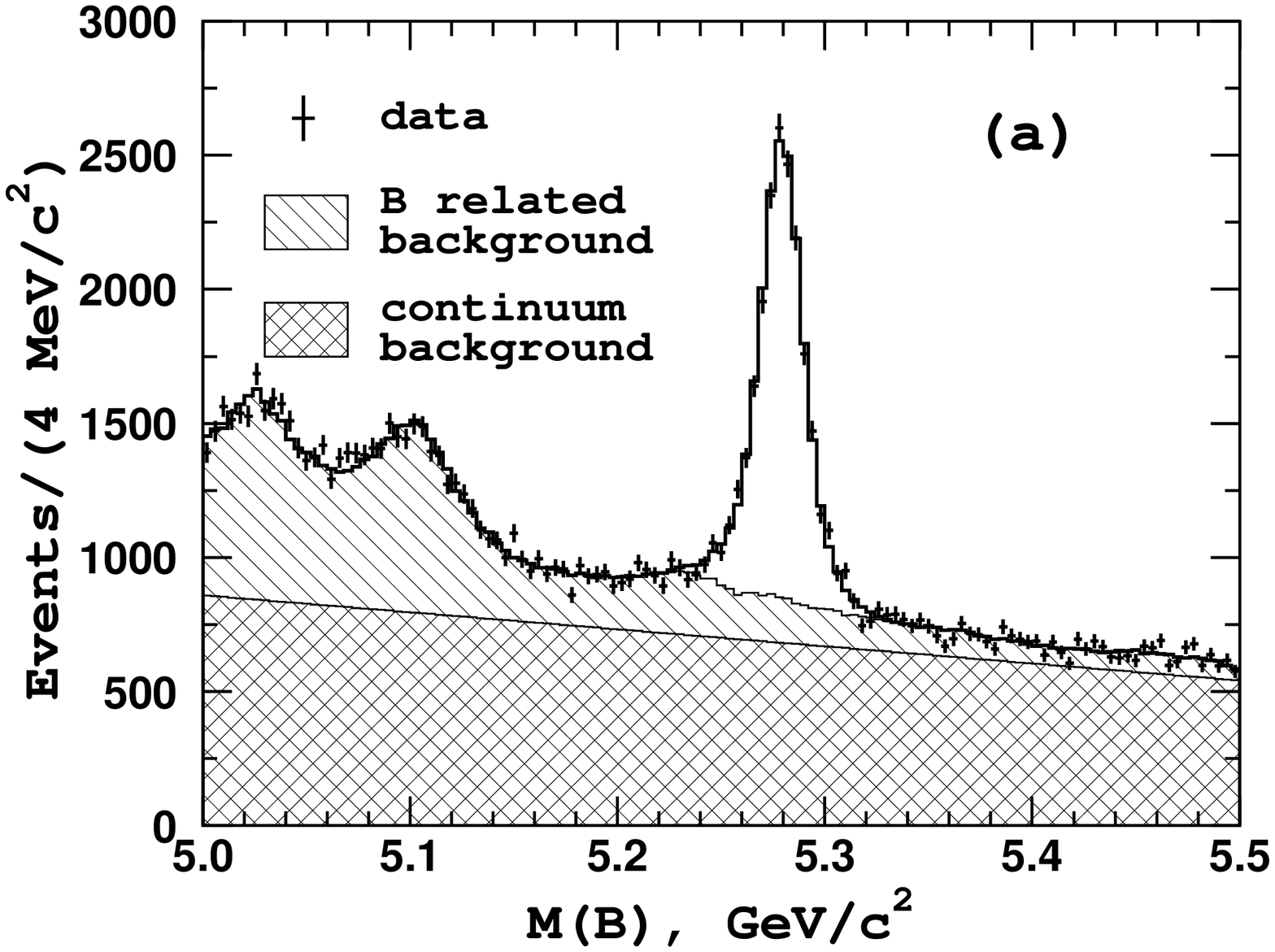} \hfill
  \hspace*{-5mm}
  \includegraphics[width=0.245\textwidth]{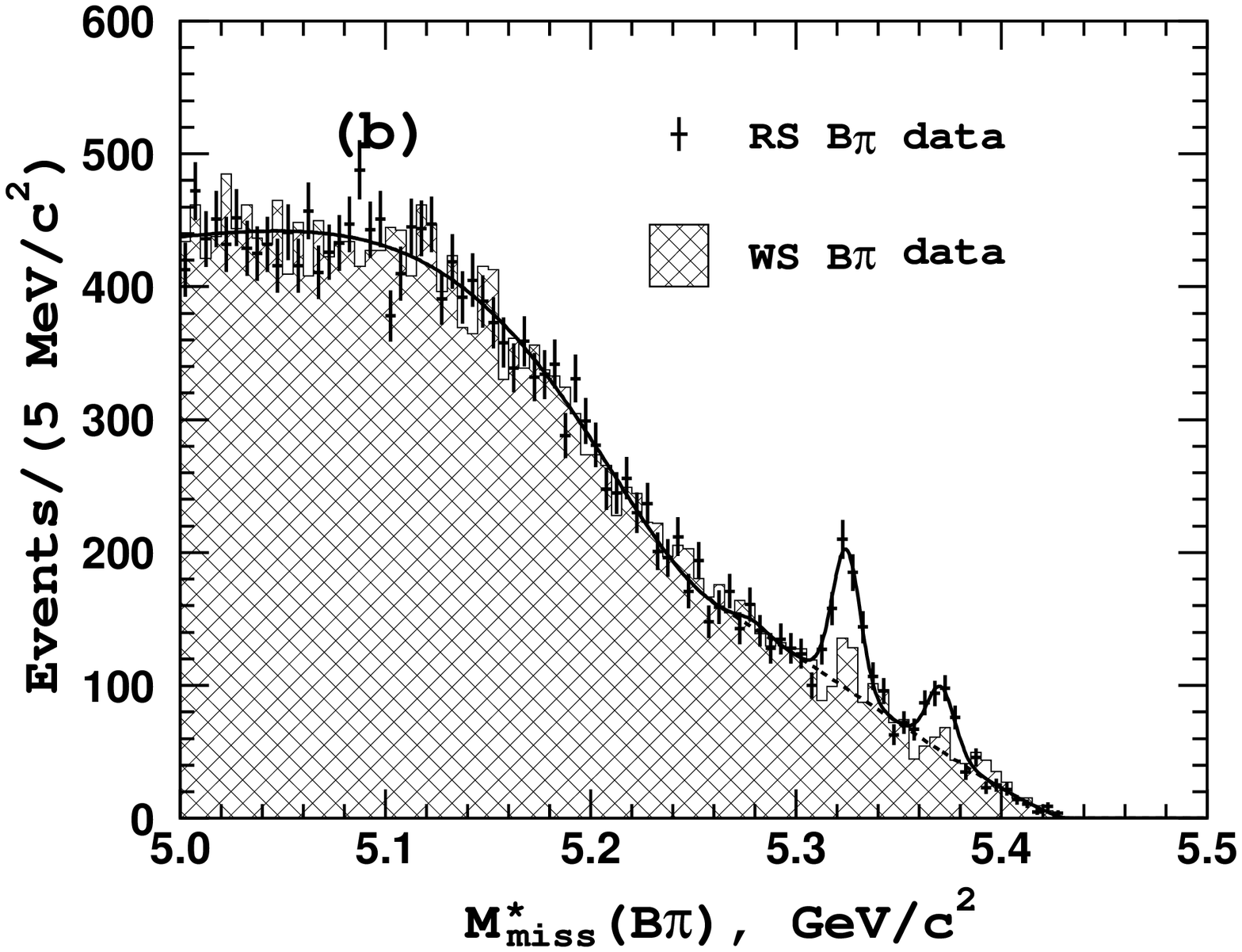}
  \caption{The (a) invariant mass and
           (b) $M^*_{\rm miss}(B\pi)$ 
           distribution for $B$ candidates in the $B$ signal region.
           Points with error bars represent the data. The open histogram
           in (a) shows the result of the fit to data. The solid line 
           in (b) shows the result of the fit to the RS $B\pi$ data; 
           the dashed line represents the background level.}
  \label{fig:zb-select}
\end{figure}

$B$ mesons are reconstructed in the following decay channels: $B^+\to
J/\psi K^{(*)+}$, $B^+\to \bar{D}^{(*)0}\pi^+$, $B^0\to J/\psi
K^{(*)0}$, $B^0\to D^{(*)-}\pi^+$ (eighteen in total).
The dominant background comes from $e^+e^-\to c\bar{c}$ continuum
events, where true $D$ mesons produced in $e^+e^-$ annihilation are
combined with random particles to form a $B$ candidate. This type of
background is suppressed using variables that characterize the event
topology.
We identify $B$ candidates by their reconstructed invariant mass $M(B)$
and momentum $P(B)$ in the center-of-mass (c.m.) frame. We require
$P(B)<1.35$~GeV/$c$ to retain $B$ mesons produced in both two-body and
multibody processes.
The $M(B)$ distribution for $B$ candidates is shown in
Fig.~\ref{fig:zb-select}(a). We perform a binned maximum likelihood fit of
the $M(B)$ distribution to the sum of a signal component parameterized
by a Gaussian function and two background components: one related to
other decay modes of $B$ mesons and one due to continuum $e^+e^-\to
q\bar{q}$ processes, where $q=u,d,s,c$.
We find $12263\pm168$ fully reconstructed $B$ mesons. The $B$
signal region is defined by requiring $M(B)$ to be within 30 to
40~MeV/$c^2$ (depending on the $B$ decay mode) of the nominal $B$ mass.

Reconstructed $B^+$ or $\bar{B}^0$ candidates are combined with
$\pi^-$'s --- the right-sign (RS) combination --- and the missing mass,
$M_{\rm miss}(B\pi)$, is calculated as $M_{\rm miss}(B\pi) =
\sqrt{(\sqrt{s}-E_{B\pi})^2/c^4-P^2_{B\pi}/c^2}$, where $E_{B\pi}$ and
$P_{B\pi}$ are the measured energy and momentum of the reconstructed
$B\pi$ combination. Signal $e^+e^-\to BB^*\pi$ events produce a narrow
peak in the $M_{\rm miss}(B\pi)$ spectrum around the nominal $B^*$ mass
while $e^+e^-\to B^{*}B^{*}\pi$ events produce a peak at $m_{B^*}+\Delta
m_{B^*}$, where $\Delta m_{B^*}=m_{B^*}-m_{B}$, due to the missed photon
from the $B^*\to B\gamma$ decay. It is important to note here that,
according to signal MC, $BB^*\pi$ events, where the reconstructed $B$ is
the one from the $B^*$, produce a peak in the $M_{\rm miss}(B\pi)$
distribution at virtually the same position as $BB^*\pi$ events, where
the reconstructed $B$ is the primary one. To remove the correlation
between $M_{\rm miss}(B\pi)$ and $M(B)$ and to improve the resolution,
we use $M_{\rm miss}^*=M_{\rm miss}(B\pi)+M(B)-m_B$ instead of $M_{\rm
  miss}(B\pi)$. The $M_{\rm miss}^*$ distribution for the RS
combinations is shown in Fig.~\ref{fig:zb-select}(b), where peaks
corresponding to the $BB^*\pi$ and $B^{*}B^{*}\pi$ signals are evident.
Combinations with $\pi^+$ --- the wrong sign (WS) combinations --- are
used to evaluate the shape of the combinatorial background. There is
also a hint for a peaking structure in the WS $M_{\rm miss}^*$
distribution, shown as a hatched histogram in
Fig.~\ref{fig:zb-select}(b). Due to $B^0-\bar{B}^0$ oscillations, we
expect a fraction of the produced $B^0$ mesons to decay as $\bar{B}^0$
given by $0.5x^2_d/(1+x^2_d)=0.1861\pm0.0024$, where $x_d$ is the $B^0$
mixing parameter~\cite{pdg}.

A binned maximum likelihood fit is performed to fit the $M_{\rm miss}^*$
distribution. ISR events produce an $M_{\rm miss}^*$ distribution
similar to that for $q\bar{q}$ events; these two components are modeled
by a single threshold function. The resolution of the signal peaks in
Fig.~\ref{fig:zb-select}(c) is dominated by the c.m.\ energy spread and
is fixed at $6.5$~MeV/$c^2$ as determined from the signal MC. The fit to
the RS spectrum yields $N_{BB\pi}=13\pm25$, $N_{BB^*\pi}=357\pm30$ and
$N_{B^{*}B^{*}\pi}=161\pm21$ signal events. The statistical significance
of the observed $BB^*\pi$ and $B^{*}B^{*}\pi$ signal is $9.3\sigma$ and
$8.1\sigma$, respectively. The statistical significance is calculated as
$\sqrt{-2\ln({\cal L}_0/{\cal L}_{\rm sig})}$, where ${\cal L}_{\rm
  sig}$ and ${\cal L}_0$ denote the likelihood values obtained with the
nominal fit and with the signal yield fixed at zero, respectively.

For the subsequent analysis, we require $|M_{\rm
  miss}^*-m_{B^*}|<15$~MeV/$c^2$ to select $BB^*\pi$ signal events and
$|M_{\rm miss}^*-(m_{B^*}+\Delta m_B)|<12$~MeV/$c^2$, where $\Delta
m_B=m_{B^*}-m_B$, to select $B^{*}B^{*}\pi$ events. For the selected
$B^*B^{(*)}\pi$ candidates, we calculate $M_{\rm miss}(\pi) =
\sqrt{(\sqrt{s}-E_\pi)^2/c^4-P^2_{\pi}/c^2}$, where $E_{\pi}$ and
$P_{\pi}$ are the reconstructed energy and momentum, respectively, of
the charged pion in the c.m.\ frame.
We perform a simultaneous binned maximum likelihood fit to the RS and WS
samples, assuming the same number and distribution of background events
in both samples and known fraction of signal events in the RS sample
that leaks to the WS sample due to mixing. To fit the $M_{\rm
  miss}(\pi)$ spectrum, we use the function
\begin{eqnarray}
F(m) = [f_{\rm sig}S(m) + B(m)]\epsilon(m)F_{\rm PHSP}(m),
\label{eq:sigfit}
\end{eqnarray}
where $m \equiv M_{\rm miss}(\pi)$; $f_{\rm sig}=1.0$
($0.1105\pm0.0016$, \cite{fsig}) for the RS (WS) sample; $S(m)$ and
$B(m)$ are the signal and background PDFs, respectively; and $F_{\rm
  PHSP}(m)$ is the phase space function. To account for the instrumental
resolution, we smear the function $F(m)$ with a Gaussian function.

\begin{figure}[h]
\includegraphics[width=0.46\textwidth]{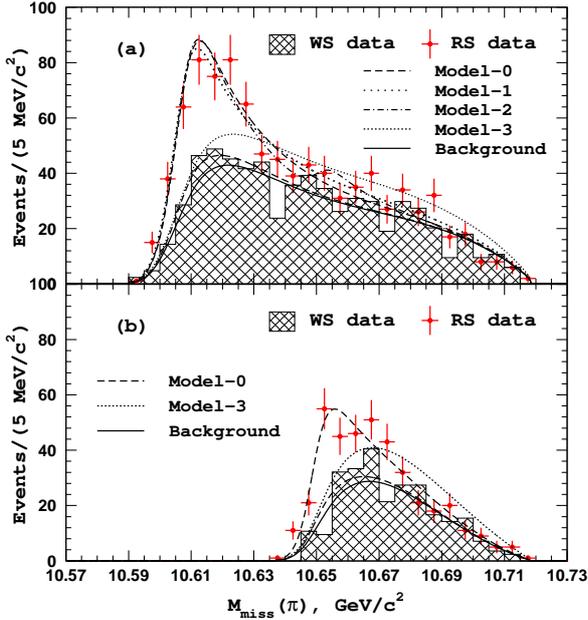} 
\hfill
  \caption{The $M_{\rm miss}(\pi)$ distribution for the 
           (a) $BB^*\pi$ and (b) $B^{*}B^{*}\pi$ candidate events.}
  \label{fig:zb-sig}
\end{figure}

We first analyze of the $BB^*\pi$ [$B^*B^*\pi$] data with the simplest
hypothesis, Model-0, that includes only the $Z_b(10610)$ [$Z_b(10650)$]
amplitude. Results of the fit are shown in Fig.~\ref{fig:zb-sig}; the
numerical results are summarized in Table~\ref{tab:zb-results}. The
fraction $f_X$ of the total three-body signal attributed to a particular
quasi-two-body intermediate state is calculated as $ f_X = {\int |{\cal
    A}_X|^2\, dm}/{\int S(m)\, dm}$, where ${\cal A}_X$ is the amplitude
for a particular component $X$ of the three-body amplitude. Next, we
extend the hypothesis to include a possible non-resonant component,
Model-1, and then the $BB^*\pi$ data is fit
to a combination of two $Z_b$ amplitudes, Model-2. In both cases,
we do not get
a statistically significant improvement in the data description: the
likelihood value is only marginally improved (see
Table~\ref{tab:zb-results}). The addition of extra components to the
amplitude also produces multiple maxima in the likelihood function. As a
result, we use Model-0 as our nominal hypothesis. Finally, we fit both
samples to a pure non-resonant amplitude (Model-3). In this case, the
fit is significantly worse.

\begin{table*}
\centering
\caption{Summary of fit results to the $M_{\rm miss}(\pi)$ distributions 
         for the three-body $BB^*\pi$ and $B^{*}B^{*}\pi$ final states.}
\medskip
\label{tab:zb-results}
  \begin{tabular}{l|c|cccccc}  \hline \hline
 Mode ~~~& Parameter & 
~Model-0~ & \multicolumn{2}{c}{Model-1} & \multicolumn{2}{c}{Model-2} & ~Model-3~ \\
      & & & Solution 1  & Solution 2 & Solution 1  & Solution 2  \\
\hline 
$BB^*\pi$
& $f_{Z_b(10610)}$      &   $1.0$   &  $1.45\pm0.24$  & $0.64\pm0.15$ & $1.01\pm0.13$  & $1.18\pm0.15$ &  $-$ \\
& $f_{Z_b(10650)}$      &   $-$     &       $-$       &      $-$      & $0.05\pm0.04$  & $0.24\pm0.11$ &  $-$\\
& $\phi_{Z_b(10650)}$, rad.   &    $-$    &       $-$       &      $-$      & $-0.26\pm0.68$ & $-1.63\pm0.14$ & $-$ \\
& $f_{\rm nr}$    &    $-$    &  $0.48\pm0.23$  & $0.41\pm0.17$ &      $-$       &       $-$      & $1.0$\\
& $\phi_{\rm nr}$, rad. &   $-$     & $-1.21\pm0.19$  & $0.95\pm0.32$ &     $-$        &      $-$      &  $-$ \\
& $-2\log{\cal{L}}$& $-304.7$ &    $-300.6$    &    $-300.5$   &    $-301.4$    &     $-301.4$  & $-344.5$ \\
\hline
$B^{*}B^{*}\pi$
& $f_{Z_b(10650)}$      &    $1.0$    & $1.04\pm0.15$ & $0.77\pm0.22$ & & &    $-$      \\
& $f_{\rm nr}$          &     $-$     & $0.02\pm0.04$ & $0.24\pm0.18$ & & &   $1.0$     \\
& $\phi_{\rm nr}$, rad. &     $-$     & $0.29\pm1.01$  & $1.10\pm0.44$  & & &   $-$      \\
& $-2\log{\cal{L}}$&  $-182.4$  &    $-182.4$    &    $-182.4$    & & &  $-209.7$    \\
\hline \hline
\end{tabular}
\end{table*}

If the parameters of the $Z_b$ resonances are allowed to float, the fit
to the $BB^*\pi$ data with Model-0 gives $10605\pm6$~MeV/$c^2$ and
$25\pm7$~MeV for the $Z_b(10610)$ mass and width, respectively, and the fit to
the $B^*B^*\pi$ data gives $10648\pm13$~MeV/$c^2$ and $23\pm8$~MeV for
the $Z_b(10650)$ mass and width, respectively. The large errors here reflect
the strong correlation between the resonance parameters.

Using the results of the fit to the $M_{\rm miss}(\pi)$ spectra with 
the nominal model (Model-0 in Table~\ref{tab:zb-results}) and the results 
of the analyses of $e^+e^-\to\Upsilon(n{\rm S})\pi^+\pi^-$~\cite{y5s2ypp} and 
$e^+e^- \to h_b(m{\rm P})\pi^+\pi^-$~\cite{y5s2hpp,note2},
we calculate the ratio of the branching fractions 
${\cal{B}}(Z_b(10610)\to B\bar{B}^*+c.c.)/{\cal{B}}(Z_b(10610)\to 
{\rm bottomonium}) = 4.76\pm0.64\pm0.75$ and 
${\cal{B}}(Z_b(10650)\to B^*\bar{B}^*)/{\cal{B}}(Z_b(10650)\to 
{\rm bottomonium}) = 2.40\pm0.44\pm0.50$.

We calculate the relative fractions for $Z_b$ decays, 
assuming that they are saturated by the already observed
$\Upsilon(n{\rm S})\pi$, $h_b(m{\rm P})\pi$, and $B^*B^{(*)}$ 
channels. The results are summarized in 
Table~\ref{tab:zb-zfracs}. 

\begin{table}
\centering
\caption{$B$ branching fractions for the $Z^+_b(10610)$ and 
         $Z^+_b(10650)$ decays. The first quoted uncertainty is 
         statistical; the second is systematic.}
\medskip
\label{tab:zb-zfracs}
  \begin{tabular}{lcc}  \hline \hline
  ~Channel~  & \multicolumn{2}{c}{Fraction, \%}   \\
              & ~$Z_b(10610)$~  & ~$Z_b(10650)$~      \\
\hline 
 $\Upsilon(1{\rm S})\pi^+$ & $0.60\pm0.17\pm0.07$ & $0.17\pm0.06\pm0.02$ \\
 $\Upsilon(2{\rm S})\pi^+$ & $4.05\pm0.81\pm0.58$ & $1.38\pm0.45\pm0.21$ \\
 $\Upsilon(3{\rm S})\pi^+$ & $2.40\pm0.58\pm0.36$ & $1.62\pm0.50\pm0.24$ \\
 $h_b(1{\rm P})\pi^+$     & $4.26\pm1.28\pm1.10$ & $9.23\pm2.88\pm2.28$ \\
 $h_b(2{\rm P})\pi^+$     & $6.08\pm2.15\pm1.63$ & $17.0\pm3.74\pm4.1$  \\
 $B^+\bar{B}^{*0}+\bar{B}^0B^{*+}$ &  $82.6\pm2.9\pm2.3$  &  $-$         \\
 $B^{*+}\bar{B}^{*0}$      &         $-$          & $70.6\pm4.9\pm4.4$  \\
\hline \hline
  \end{tabular}
\end{table}

In conclusion, we report the first observations of the three-body
$e^+e^- \to BB^*\pi$ and $e^+e^-\to B^{*}B^{*}\pi$ processes with a
statistical significance above $8\sigma$. The analysis of the
$B^{(*)}B^*$ mass spectra indicates that the total three-body rates are
dominated by the intermediate $e^+e^- \to Z_b(10610)^\mp\pi^\pm$ and
$e^+e^- \to Z_b(10650)^\mp\pi^\pm$ transitions for the $BB^*\pi$ and
$B^{*}B^{*}\pi$ final states, respectively.

\section{Observation of the decay $\Lambda^{+}_{c} \rightarrow p K^{+} \pi^{-}$}
\label{sec-5-lamcpkpi}

Several doubly Cabibbo-suppressed (DCS) decays of charmed mesons have
been observed. Their measured branching ratios with respect to
corresponding Cabibbo-favored (CF) decays play an important role in
constraining models of the decay of charmed hadrons and in the study of
flavor-$SU(3)$ symmetry~\cite{dcs_mesons}. Because of the smaller
production cross sections for charmed baryons, DCS decays of charmed
baryons have not yet been observed, and only an upper limit,
$\mathcal{B}(\Lambda^{+}_{c} \rightarrow p K^{+}
\pi^{-})/\mathcal{B}(\Lambda^{+}_{c} \rightarrow p K^{-}
\pi^{+})<0.46\%$ with 90\% confidence level, has been reported by the
FOCUS Collaboration~\cite{dcs_focus}. Theoretical calculations of DCS
decays of charmed baryons have been limited to two-body decay
modes~\cite{dcs_theory1, dcs_theory2}.

Recently Belle reported the first observation of the DCS decay
$\Lambda^{+}_{c} \rightarrow p K^{+} \pi^{-}$ and the measurement of its
branching ratio with respect to its counterpart CF decay
$\Lambda^{+}_{c} \rightarrow p K^{-} \pi^{+}$. In the
letter~\cite{lamcpkpi}, Belle reports the first observation of the DCS
decay $\Lambda^{+}_{c} \rightarrow p K^{+} \pi^{-}$ and the measurement
of its branching ratio with respect to its counterpart CF decay
$\Lambda^{+}_{c} \rightarrow p K^{-} \pi^{+}$. Unlike charmed meson
decays, internal $W$ emission and $W$ exchange are not suppressed for
charmed baryon decays. In previous studies of CF or singly
Cabibbo-suppressed (SCS) decays of $\Lambda_c^{+}$ and $\Xi_{c}^{0}$,
direct evidence of $W$ exchange and internal $W$ emission has been
observed~\cite{scs}. When we consider that $W$
exchange is prohibited in $\Lambda^{+}_{c} \rightarrow p K^{+} \pi^{-}$
but allowed in $\Lambda^{+}_{c} \rightarrow p K^{-} \pi^{+}$, the
contribution of $W$ exchange to $\Lambda^{+}_{c}$ decays can be
estimated by comparing the measured branching ratio with the na\"{\i}ve
expectation~\cite{dcs_focus}, $\tan^{4}{\theta_{{\mathrm c}}}$
(0.285$\%$), where $\theta_{{\mathrm c}}$ is the Cabibbo mixing
angle~\cite{cabibbo_angle} and $\sin{\theta_{{\mathrm
      c}}}=0.225\pm0.001$~\cite{pdg}. This approach does not take into
account effects of flavor-$SU(3)$ symmetry breaking.

A $\Lambda_{c}^{+}$ candidate is reconstructed from the three charged
hadrons. To suppress combinatorial backgrounds, especially from $B$
meson decays, we place a requirement on the scaled momentum:
$x_{p}>0.53$, where $x_{p}$ is defined as $p^{*}/\sqrt{E_{\rm
    cm}^{2}/4-M^{2}}$; here, $E_{\rm cm}$ is the total center-of-mass
energy, $p^{*}$ is the momentum in the center-of-mass frame, and $M$ is
the mass of the $\Lambda_{c}^{+}$ candidate. In addition, the $\chi^{2}$
value from the common vertex fit of the charged tracks must be less than
40.



\begin{figure}[h]
  \includegraphics[width=0.245\textwidth]{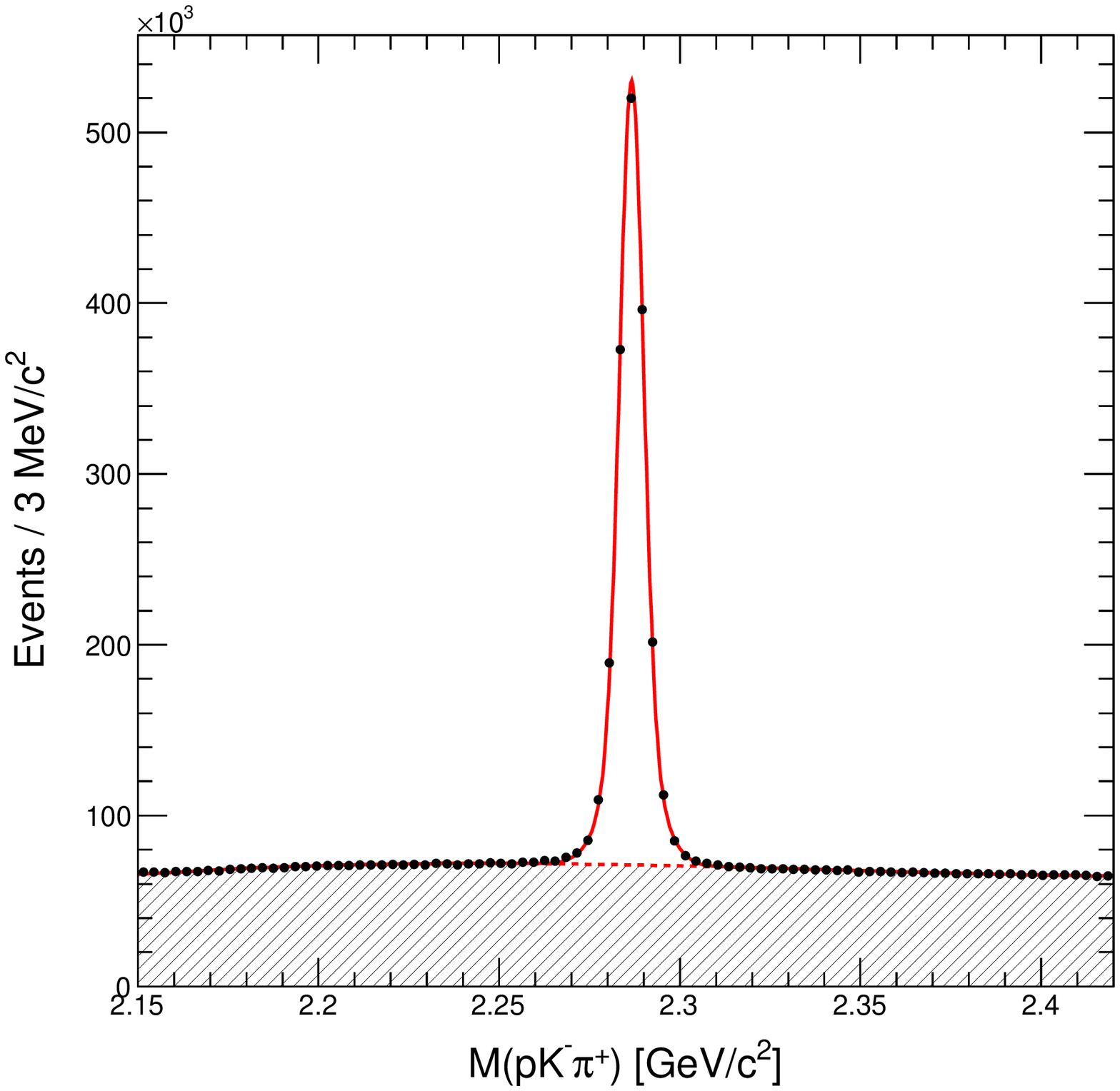} \hfill
  \hspace*{-5mm}
  \includegraphics[width=0.245\textwidth]{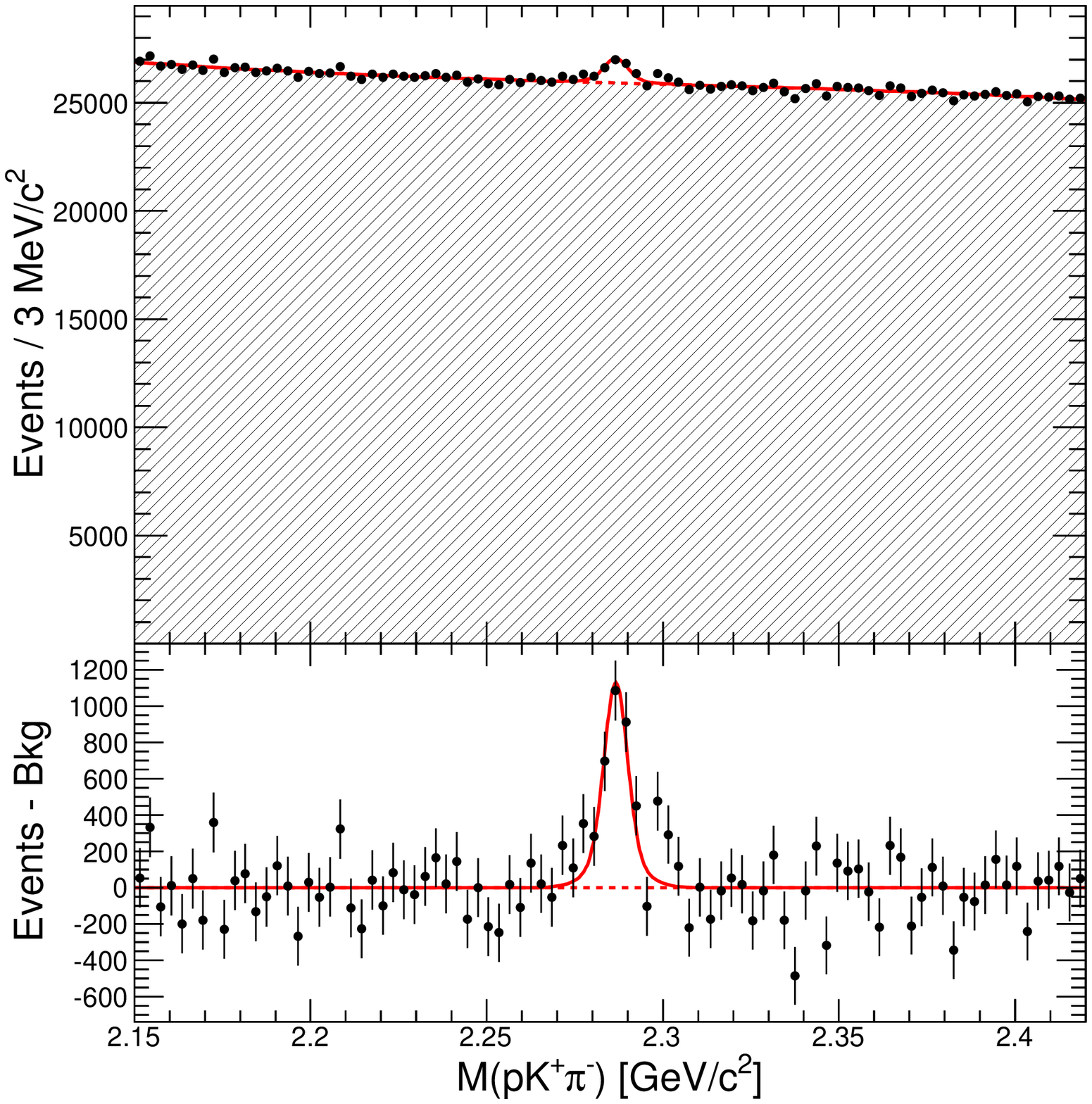}
  \caption{Distribution of $M(pK^{-}\pi^{+})$ (left), $M(pK^{+}\pi^{-})$
    (right top) and residuals of data with respect to the fitted
    combinatorial background (right bottom). The curves indicate the fit
    result: the full fit model (solid) and the combinatoric background
    only (dashed).}
  \label{fig:lamc}
\end{figure}

Figures~\ref{fig:lamc}, left and \ref{fig:lamc}, right show invariant mass
distributions, $M(pK^{-}\pi^{+})$ (CF) and $M (pK^{+}\pi^{-})$ (DCS), respectively,
with the final selection criteria. DCS decay events are clearly observed
in $M(pK ^{+}\pi^{-})$. We perform a binned least-$\chi^{2}$ fit to the
two distributions from 2.15~$\mathrm{GeV/}c^{2}$ to
2.42~$\mathrm{GeV/}c^{2}$ with 0.01 $\mathrm{MeV/}c^{2}$ bin width, and
the figures are drawn with merged bins.
The DCS decay has a peaking background from the SCS decay
$\Lambda_{c}^{+}\rightarrow\Lambda K^{+}$ with $\Lambda \rightarrow
p\pi^{-}$, which has the same final state topology. However, because of
the long $\Lambda $ lifetime, many of the $\Lambda $ vertexes are
displaced by several centimeters from the main vertex so the 
geometrical
requirements suppress most of this background. The remaining
SCS-decay yield is included in the signal yield of $\Lambda^{+}_{c}
\rightarrow p K^{+} \pi^{-}$ decay and is estimated via the relation
\begin{multline}
\label{eq:scs_signal_yield}
 \mathcal{N}(SCS;\Lambda \rightarrow p\pi^{-})= \\
\frac {\mathcal{\epsilon} (SCS;\Lambda \rightarrow p\pi^{-})}{\mathcal{\epsilon}(CF)} \frac {\mathcal{B}(SCS;\Lambda \rightarrow p\pi^{-})}{\mathcal{B}(CF)} \mathcal{N}(CF), 
\end{multline}
where $\mathcal{N}(CF)$ is the signal yield of the CF decay,
$\mathcal{B}(SCS;\Lambda \rightarrow
p\pi^{-})/\mathcal{B}(CF)=(0.61\pm0.13)\%$ is the branching
ratio~\cite{pdg}, and $\mathcal{\epsilon} (SCS;\Lambda \rightarrow
p\pi^{-})/\mathcal{\epsilon}(CF)=0.023$ is the relative efficiency found
using MC samples.
After subtraction of this SCS component, the
signal yield of the DCS decay is $3379\pm380\pm78$, where the first
uncertainty is statistical and the second is systematic due to this
subtraction.
To estimate the statistical significance of the DCS signal, we exclude
the SCS signal by vetoing events with
$1.1127~\mathrm{GeV/}c^{2}<M(p\pi^{-})<1.1187~\mathrm{GeV/}c^{2}$. The
significance is estimated as
$\sqrt{-2\ln{(\mathcal{L}_{0}/\mathcal{L})}}$, where $\mathcal{L}_{0}$
and $\mathcal{L}$ are the maximum likelihood values from binned maximum
likelihood fits with the signal yield fixed to zero and allowed to
float, respectively. The calculated significance corresponds to
9.4$\sigma$.

The branching ratio, $\mathcal{B}(\Lambda^{+}_{c} \rightarrow p K^{+}
\pi^{-})/\mathcal{B}(\Lambda^{+}_{c} \rightarrow p K^{-} \pi^{+})$, is
$(2.35\pm0.27\pm0.21)\times 10^{-3}$, where the uncertainties are
statistical and systematic, respectively. The branching fraction of the
CF decay, $(6.84\pm0.24^{+0.21}_{-0.27})\times 10^{-2}$, was already
well-measured in a previous Belle analysis~\cite{br_cf}. Combining that
with our measurement, we determine the absolute branching fraction of
the DCS decay to $(1.61\pm0.23^{+0.07}_{-0.08})\times 10^{-4}$, where
the first uncertainty is the total uncertainty of the branching ratio
and the second is uncertainty of the branching fraction of CF
decay. This measured branching ratio corresponds to $(0.82 \pm
0.12)\tan^{4}{\theta_{{\mathrm c}}}$, where the uncertainty is the
total.

%
%

\end{document}